\documentclass[aps,pre,twocolumn,groupedaddress]{revtex4-1}
\usepackage{amsfonts,amssymb,amsmath} 
\usepackage{color} 
\usepackage{graphicx} 
\usepackage{epstopdf,mathrsfs} 
\usepackage[colorlinks,linktocpage,linkcolor=blue]{hyperref}

\usepackage{xcolor}

\begin{document}

\title{Ergodic property of Langevin systems with superstatistical, uncorrelated or correlated diffusivity}

\author{Xudong Wang$^1$}
\email{xdwang14@njust.edu.cn}
\author{Yao Chen$^2$}
\affiliation{$^1$School of Science, Nanjing University of Science and Technology, Nanjing, 210094, P.R. China \\
$^2$School of Mathematics and Statistics, Gansu Key Laboratory
of Applied Mathematics and Complex Systems, Lanzhou University, Lanzhou 730000,
P.R. China}



\begin{abstract}
Brownian yet non-Gaussian diffusion has recently been observed in numerous biological and active matter system. The cause of the non-Gaussian distribution have been elaborately studied in the idea of a superstatistical dynamics or a diffusing diffusivity. Based on a random diffusivity model, we here focus on the ergodic property and the scatter of the amplitude of time-averaged mean-squared displacement (TAMSD). Further, we individually investigate this model with three categories of diffusivities, including diffusivity being a random variable $D$, a time-dependent but uncorrelated diffusivity $D(t)$, and a correlated stochastic process $D(t)$. We find that ensemble-averaged TAMSDs are always normal while ensemble-averaged mean-squared displacement can be anomalous. Further, the scatter of dimensionless amplitude is determined by the time average of diffusivity $D(t)$. Our results are valid for arbitrary diffusivities.

\end{abstract}

\maketitle

\section{Introduction}

Since the Robert Brown's experiments of observing the erratic motion of granules extracted from pollen grains suspending in water \cite{Brown:1828}, Brownian motion has attracted great attention of numerous scientists to improve the experiments \cite{Nordlund:1914} and to develop the theory of diffusion \cite{Einstein:1905,Sutherland:1905,Smoluchowski:1906,Langevin:1908}. Even when a variety of anomalous diffusion processes characterized by nonlinearly growing ensemble-averaged mean-squared displacement (EAMSD) are observed \cite{BouchaudGeorges:1990,MetzlerKlafter:2000,MetzlerKlafter:2004,MetzlerJeonCherstvyBarkai:2014,HoflingFranosch:2013,Norregaard-etal:2017} by use of the advance of modern microscopy techniques and massive progress in supercomputing, the research of Brownian motion has never stopped.

As we all know, Brownian motion has three fundamental properties \cite{VanKampen:1992,CoffeyKalmykovWaldron:2004}. The first one is its normal diffusion behavior described by a linearly growing EAMSD in time:
\begin{equation}\label{LinearMSD}
  \langle x^2(t)\rangle = 2Dt.
\end{equation}
Second, the displacement of Brownian motion has a Gaussian shape of probability density function (PDF):
\begin{equation}\label{Gaussian}
  G(x,t|D)=\frac{1}{\sqrt{4\pi Dt}}\exp\left(-\frac{x^2}{4Dt}\right).
\end{equation}
The last one is that Brownian motion has independent increments. The absence of independent increments is usually the essential reason of displaying a anomalous diffusion, such as continuous-time random walk (CTRW) or L\'{e}vy walk with correlated waiting time or jump length \cite{ChechkinHofmannSokolov:2009,ChenWangDeng:2019,TejedorMetzler:2010,MagdziarzMetzlerSzczotkaZebrowski:2012-1,MagdziarzMetzlerSzczotkaZebrowski:2012-2},
and the class of viscoelastic diffusion described by the generalized Langevin equation with power-law friction kernel \cite{Lutz:2001,Goychuk:2012,SlezakMetzlerMagdziarz:2018,DengBarkai:2009} and fractional Brownian motion \cite{MandelbrotNess:1968,MeerschaertSabzikar:2013,ChenWangDeng:2017}.

A new class of diffusive dynamics has recently been observed in a large range of complex systems, which is named as Brownian yet non-Gaussian process due to the coexisting phenomenon of linear EAMSD and non-Gaussian PDF. Examples include polystyrene beads diffusing on the surface of lipid tubes \cite{WangAnthonyBaeGranick:2009} or in networks \cite{WangAnthonyBaeGranick:2009,ToyotaHeadSchmidtMizuno:2011,SilvaStuhrmannBetzKoenderink:2014}, as well as the diffusion of tracer molecules on polymer thin films \cite{Bhattacharya-etal:2013} and in simulations of a two dimensional discs \cite{KimKimSung:2013}. Instead of the Gaussian shape, the PDF of this new class of processes is characterized by an exponential distribution
\begin{equation}\label{Exponential}
  p(x,t)\simeq \frac{1}{2\sqrt{D_0t}}\exp\left(-\frac{|x|}{\sqrt{D_0t}}\right),
\end{equation}
where $D_0$ is the effective diffusivity. The physical interpretation of the non-Gaussian diffusion was given by a superstatistical approach \cite{Beck:2001,BeckCohen:2003,Beck:2006}. More specifically, each particle undergoes a Brownian diffusion with its own diffusion coefficient $D$ which does not change considerably in a short time less than the characteristic time of correlation in a system. By employing an exponential distribution of the diffusion coefficient
\begin{equation}\label{Exponential-D}
  \pi(D)=\frac{1}{D_0}\exp\left(-\frac{D}{D_0}\right)
\end{equation}
with the mean $D_0$, and averaging the Gaussian PDF in Eq. \eqref{Gaussian} for a single diffusion coefficient $D$ over the exponential distribution in Eq. \eqref{Exponential-D}, i.e.,
\begin{equation}\label{PDF-x}
\begin{split}
    p(x,t)&=\int_0^\infty \pi(D)G(x,t|D)dD  \\
    &=\frac{1}{\sqrt{4D_0t}}\exp\left(-\frac{|x|}{\sqrt{D_0t}}\right),
\end{split}
\end{equation}
presenting the non-Gaussian-shaped PDF as Eq. \eqref{Exponential} appears \cite{WangKuoBaeGranick:2012,HapcaCrawfordYoung:2009,ChubynskySlater:2014,ChechkinSenoMetzlerSokolov:2017}.

Most of the existing works mainly focus on the PDF of the Brownian yet non-Gaussian process. To describe the crossover from the exponential distribution at short time to the Gaussian distribution at long time, Chubynsky and Slater proposed a diffusing diffusivity model with diffusivity undergoing a random walk \cite{ChubynskySlater:2014}, and Chechkin {\it et al.} established a minimal model for processes with diffusing diffusivity under the framework of Langevin equation \cite{ChechkinSenoMetzlerSokolov:2017}. To further describe the stochastic particle motions in complex environments, the idea of random diffusivity has been applied to generalized grey Brownian motion \cite{SposiniChechkinSenoPagniniMetzler:2018}, generalized Langevin equation \cite{SlezakMetzlerMagdziarz:2018}, fractional Brownian motion \cite{JainSebastian:2018,MackalaMagdziarz:2019}, and the corresponding first-passage statistics \cite{SposiniChechkinMetzler:2019}, and so on.

In this paper, we pay more attention to the (non)ergodic behavior of the Langevin system with diffusing diffusivity, especially on the scatter of the amplitude of time-averaged mean-squared displacement (TAMSD), which is defined as \cite{MetzlerJeonCherstvyBarkai:2014,BurovJeonMetzlerBarkai:2011}
\begin{equation}\label{Def-TAMSD}
  \overline{\delta^2(\Delta)}=\frac{1}{T-\Delta}\int_{0}^{T-\Delta} [x(t+\Delta)-x(t)]^2 dt.
\end{equation}
Here, it is usually assumed that the lag time $\Delta$ is much smaller than the total measurement time $T$ for obtaining a good statistical property. Based on the advance of single-particle tracking techniques, scientists often evaluate the recorded time series in terms of TAMSD to study the diffusion behavior of particles in living cell \cite{GoldingCox:2006,WeberSpakowitzTheriot:2010,BronsteinIsraelKeptenMaiTalBarkaiGarini:2009}.
A process is called ergodic if the TAMSD and EAMSD are equivalent, i.e., $\overline{\delta^2(\Delta)}=\langle x^2(\Delta)\rangle$ as $T\rightarrow\infty$, such as Brownian motion and (tempered) fractional Brownian motion \cite{Goychuk:2012,DengBarkai:2009,ChenWangDeng:2017}.
The ergodic properties has been discussed to a certain extent for a stochastic Langevin equation with the local diffusivity fluctuating in time \cite{CherstvyMetzler:2016}. But the specific distribution of the scatter $\phi(\eta)$ is lacking, where the dimensionless random variable $\eta$ is defined as \cite{MetzlerJeonCherstvyBarkai:2014,BurovJeonMetzlerBarkai:2011}
\begin{equation}\label{Def-eta}
  \eta:=\frac{\overline{\delta^2(\Delta)}}{\langle \overline{\delta^2(\Delta)}\rangle}.
\end{equation}
The statistics of the scatter of the dimensionless amplitude $\eta$ for a set of individual trajectories at a given lag time $\Delta$ is a useful indicator to classify the numerous anomalous diffusion processes.

As mentioned above, the diffusivity in the existing works can be divided into three categories. The simplest one is that the diffusivity is a time-independent random variable $D$, corresponding to the superstatistical approach \cite{WangAnthonyBaeGranick:2009,WangKuoBaeGranick:2012,SposiniChechkinSenoPagniniMetzler:2018,SlezakMetzlerMagdziarz:2018}. The second one is a stochastic diffusivity $D(t)$ which is time-dependent, but uncorrelated between different times \cite{CherstvyMetzler:2016}. The last categories is that diffusivity is an autocorrelated stochastic process, namely, diffusing diffusivity \cite{ChubynskySlater:2014,ChechkinSenoMetzlerSokolov:2017}.
The ergodic properties, the distribution of the scatter of the dimensionless amplitude, as well as the ergodicity breaking (EB) parameter will be discussed both theoretically and numerically in detail in this paper.

The structure of this paper is as follows. In Sec. \ref{Sec2}, the random diffusivity model, together with three kinds of diffusivities, are introduced. Based on this model, we evaluate the corresponding EAMSD and TAMSD in Sec. \ref{Sec3}, and the scatter of TAMSDs for three different cases in Sec. \ref{Sec4}.  Some discussions and summaries are provided in Sec. \ref{Sec5}. For convenience, we put some mathematical details in Appendix.

\section{Random diffusivity model}\label{Sec2}
We consider the following one-dimensional overdamped Langevin equation
\begin{equation}\label{DDmodel}
  \frac{d}{dt}x(t)=\sqrt{2D(t)}\xi(t),
\end{equation}
where $x(t)$ denotes particle's trajectory and $\xi(t)$ is a Gaussian white noise with zero mean $\langle\xi(t)\rangle=0$ and the correlation function $\langle\xi(t_1)\xi(t_2)\rangle=\delta(t_1-t_2)$. In this random diffusivity model, the diffusivity $D(t)$ can be a random variable or a diffusion process. A deterministic diffusivity $D_0$ (i.e., $D(t)\equiv D_0$) make this system \eqref{DDmodel} return back to the classical Brownian motion with a linear growth MSD and a Gaussian-shaped PDF in Eqs. \eqref{LinearMSD} and \eqref{Gaussian}, respectively.

As a representative example in the first scenario that diffusivity $D$ is a random variable, the exponential distribution of $D$ in Eq. \eqref{Exponential-D} results in the well-known Brownian non-Gaussian diffusion.
For the second scenario with a time-dependent but uncorrelated diffusivity $D(t)$, one common example is
the exponential distribution
\begin{equation}\label{time-dependet-dist}
  \pi(D,t)=\frac{1}{f(t)}\exp\left(-\frac{D}{f(t)}\right)
\end{equation}
with a positive time-dependent mean $f(t)$. Specifically, the Rayleigh distribution and the stretched Gaussian distribution can be considered in this case.
Note that the different for of distribution $\pi(D,t)$ of time-dependent diffusivity $D(t)$ only change the shape of the PDF $p(x,t)$. The diffusion behaviors, normal or anomalous, are determined by the scaling relation between spatial and temporal variables in the PDF $\pi(D,t)$ of diffusivity, rather than the specific scaling functions (e.g., exponential function in Eq. \eqref{time-dependet-dist}).
Since the PDF is not the key point in this paper, we only take the exponential distribution in Eq. \eqref{time-dependet-dist} as the example of a time-dependent but uncorrelated diffusivity $D(t)$.

In the last scenario with a diffusing diffusivity, we mainly consider three representative examples: diffusivity is the square of Brownian motion \cite{Li:1992,Majumdar:2005,Maccone:2012} ($D(t)=B^2(t)$); diffusivity is the reflected Brownian motion \cite{Grebenkov:2007,Grebenkov:2007-2,BlanchetChen:2015} ($D(t)=|B(t)|$); diffusivity is the square of Ornstein-Uhlenbeck (OU) process $D(t)=Y^2(t)$ where $\dot{Y}(t)=-Y(t)+\xi(t)$ \cite{ChechkinSenoMetzlerSokolov:2017,Dankel:1991,CheriditoKawaguchiMaejima:2003,WangChenDeng:2019}. Though we take the three examples above where $D(t)$ is positive, our results of this paper are valid for diffusivity being any diffusion process.

\section{EAMSD and TAMSD}\label{Sec3}
Note that Eq. \eqref{PDF-x} is only valid when diffusivity is a time-independent random variable. For a time-dependent diffusivity $D(t)$, we can directly perform integral on Eq. \eqref{DDmodel} to obtain the trajectory $x(t)$ and further take ensemble average to obtain the EAMSD, i.e.,
\begin{equation}\label{EAMSD}
  \begin{split}
    \langle x^2(t)\rangle
    &=2\int_0^tdt_1'\int_0^tdt_2'\left\langle \sqrt{D(t_1')D(t_2')}\xi(t_1')\xi(t_2')\right\rangle \\
    &=2\int_0^tdt_1'\int_0^tdt_2'\left\langle \sqrt{D(t_1')D(t_2')}\right\rangle \delta(t_1'-t_2') \\
    &=2\int_0^t  \langle D(t')\rangle dt',
  \end{split}
\end{equation}
where we have used the independence between diffusivity $D(t)$ and the white noise $\xi(t)$ and $\langle\xi(t_1)\xi(t_2)\rangle=\delta(t_1-t_2)$ in the second line.
Equation \eqref{EAMSD} implies that the EAMSD only depends on the mean value of diffusivity $D(t)$. An increasing trend of mean $\langle D(t)\rangle$ leads to superdiffusion behavior while a decreasing one yields subdiffusion behavior. The normal diffusion is recovered when diffusivity is a time-independent random variable or the mean $\langle D(t)\rangle$ tends to a constant for long time.

With similar procedure to Eq. \eqref{EAMSD}, and using the property of $\delta$-correlation of white noise $\xi(t)$, the autocorrelation function of $x(t)$ satisfies
\begin{equation}\label{ACF-x}
  \langle x(t)x(t+\Delta)\rangle  =  \langle x^2(t)\rangle =2\int_0^t  \langle D(t')\rangle dt'.
\end{equation}
Then we substitute Eq. \eqref{ACF-x} into the definition of TAMSD in Eq. \eqref{Def-TAMSD}, and thus obtain the ensemble-averaged TAMSD
\begin{equation}\label{EATAMSD}
  \begin{split}
    \langle\overline{\delta^2(\Delta)}\rangle
    &=\frac{1}{T-\Delta}\int_{0}^{T-\Delta} \langle x^2(t+\Delta)\rangle-\langle x^2(t)\rangle dt   \\
    &=\frac{2}{T-\Delta}\int_{0}^{T-\Delta} \int_t^{t+\Delta} \langle D(t')\rangle dt' dt,
  \end{split}
\end{equation}
which only depends on $\langle D(t)\rangle$ as the EAMSD. In particular, single-point information of diffusivity $D(t)$ is sufficient here, which is different from two-point approach for models with power-law distributed waiting times, like subdiffusive CTRW \cite{LubelskiSokolovKlafter:2008,HeBurovMetzlerBarkai:2008,BurovMetzlerBarkai:2010}, L\'{e}vy walk \cite{FroembergBarkai:2013,FroembergBarkai:2013-2,GodecMetzler:2013}, and Langevin dynamics coupled with an inverse subordinator \cite{BauleFriedrich:2005,BauleFriedrich:2007,WangChenDeng:2019,ChenWangDeng:2019-3,ChenWangDeng:2019-2}.

Based on this observation Eq. \eqref{EATAMSD} and the priori condition $\Delta\ll T$, the ensemble-averaged TAMSD have the asymptotic behavior
\begin{equation}\label{EATAMSD-Asym}
  \langle\overline{\delta^2(\Delta)}\rangle
   \simeq \frac{2\Delta}{T}\int_0^T \langle D(t')\rangle dt'
   =2\Delta\langle \overline{D(t)}\rangle,
\end{equation}
where $\overline{D(t)}$ is the time average of diffusivity $D(t)$ throughout the entire observation period $[0,T]$. Since the time variable of quantity $\langle \overline{D(t)}\rangle$ in Eq. \eqref{EATAMSD-Asym} is total measurement time $T$, rather than lag time $\Delta$, the TAMSD grows linearly on $\Delta$ and presents an eternal normal diffusion.
This phenomenon makes the TAMSD deviate from the EAMSD which is anomalous only if the mean value $\langle \overline{D(t)}\rangle$ is time-dependent for long time. Therefore, a constantly time-dependent diffusivity $D(t)$ leads to a weak nonergodic behavior.

\section{Scatter of TAMSD}\label{Sec4}

In general, the TAMSD is a stochastic process due to the randomness of the integrand $[x(t+\Delta)-x(t)]^2$ in Eq. \eqref{Def-TAMSD}. The ensemble averages on TAMSD in Eq. \eqref{EATAMSD-Asym} miss the information of randomness of TAMSD. Therefore, as a more detailed quantity, the scatter of TAMSD is a useful indicator to distinguish various anomalous diffusion processes. This quantity will be investigated on random diffusivity model in this section.

For subdiffusive CTRW with power-law distributed waiting times, it can be observed quite often for a wide range of $t$ that no jump event happens between time $t$ and $t+\Delta$, which leads to $[x(t+\Delta)-x(t)]^2=0$ for many different values of $t$. By contrast, the jumps occur all the time with a certain frequency in the diffusing diffusivity model Eq. \eqref{DDmodel}.
A constant diffusivity $D$ of classical Brownian motion indicates that the magnitude of jumps at each step are of the same size, whereas the diffusing diffusivity resembles the inhomogeneous magnitude of different jumps.
From another point of view, the discrepant magnitude of jumps can be regarded as varying numbers of jumps with the same magnitude and the variance of each jump is a finite constant. In this sense, the number of jumps between time $t$ and $t+\Delta$ can be written as
\begin{equation}
  2\int_t^{t+\Delta} D(t')dt',
\end{equation}
which is random due to the randomness of diffusivity $D(t)$. Combining it with the definition of TAMSD in Eq. \eqref{Def-TAMSD}, we find that the TAMSD behaves as
\begin{equation}\label{Dist-TAMSD}
  \overline{\delta^2(\Delta)}\simeq C\int_0^{T-\Delta} \int_t^{t+\Delta} D(t')dt' dt,
\end{equation}
where $C$ is a constant independent of diffusivity $D$. Constant $C$ can be determined by taking ensemble average on both sides of Eq. \eqref{Dist-TAMSD}, which leads to $C=2/(T-\Delta)$ by use of Eq. \eqref{EATAMSD}. Now we consider the condition $\Delta\ll T$ and the long time $T$ limit, similar to Eq. \eqref{EATAMSD-Asym}, we have a simpler form for TAMSD, i.e.,
\begin{equation}\label{TAMSD}
  \overline{\delta^2(\Delta)}\simeq \frac{2\Delta}{T}\int_0^T D(t)dt=2\Delta\overline{D(t)}.
\end{equation}
This result is consistent with Eq. \eqref{EATAMSD-Asym} by taking ensemble average on both sides.

Now we can evaluate the TAMSD for different diffusivity $D(t)$ based on the general result in Eq. \eqref{TAMSD}. For many anomalous diffusion processes, the TAMSD is a random variable and shows pronounced trajectory-to-trajectory variations, such as L\'{e}vy walk \cite{FroembergBarkai:2013,FroembergBarkai:2013-2,GodecMetzler:2013}, L\'{e}vy flight \cite{VahabiSchulzShokriMetzler:2013,FroembergBarkai:2013-2,BurneckiWeron:2010}, quenched models \cite{Massignan-etal:2014,MiyaguchiAkimoto:2011}, heterogeneous diffusion processes \cite{CherstvyChechkinMetzler:2013,CherstvyMetzler:2014,WangDengChen:2019,LeibovichBarkai:2019} and so on. The stochasticity of TAMSD can be measured by the scatter PDF $\phi(\eta)$ where the dimensionless random variable is equal to
\begin{equation}\label{Def-eta}
  \eta=\frac{\overline{\delta^2(\Delta)}}{\langle \overline{\delta^2(\Delta)}\rangle}\simeq\frac{\overline{D(t)}}{\langle \overline{D(t)}\rangle}
\end{equation}
for large $T$, where we have used the relation in Eq. \eqref{TAMSD}. Here we do not use the notation $\eta(\Delta)$ since it is independent on lag time $\Delta$. This result is universal for the random diffusivity model Eq. \eqref{DDmodel}. It holds that $\phi(\eta)=\delta(\eta-1)$ for an ergodic process, while a nonergodic process shows a broad distribution of $\eta$. A measure of the scatter of TAMSD is the variance of dimensionless random variable $\eta$, which is also named as ergodicity breaking (EB) parameter:
\begin{equation}\label{Def-EB}
\begin{split}
    \textrm{EB}=\langle \eta^2\rangle-\langle\eta\rangle^2
    =\frac{\langle(\overline{\delta^2(\Delta)})^2\rangle
    -\langle\overline{\delta^2(\Delta)}\rangle^2}{\langle\overline{\delta^2(\Delta)}\rangle^2}.
\end{split}
\end{equation}
The EB parameter for the diffusing diffusivity model Eq. \eqref{DDmodel} can be obtained by substituting Eq. \eqref{Def-eta} into Eq. \eqref{Def-EB}, i.e.,
\begin{equation}\label{EB-Dt}
  \textrm{EB}\simeq\frac{\langle(\overline{D(t)})^2\rangle
    -\langle\overline{D(t)}\rangle^2}{\langle\overline{D(t)}\rangle^2}
\end{equation}
under the condition $\Delta\ll T$.
If the EB parameter tends to zero as $T\rightarrow\infty$, then $\eta$ converges to its mean $\langle\eta\rangle=1$ in the mean square sense.

For the three categories of diffusivity $D(t)$ mentioned in Sec. \ref{Sec2}, the scatter of the amplitude of the dimensionless variable $\eta$ in Eq. \eqref{Def-eta} and the corresponding EB parameter in Eq. \eqref{EB-Dt} will be explicitly discussed in the following.

\subsection{Superstatistical approach}
The first kind of diffusivity is a time-independent random variable $D(t)\equiv D$, corresponding to the superstatistical phenomenon. A typical example is the Brownian diffusion or the Ornstein-Uhlenbeck process with two phase ($D$ alternates between two different positive values, a larger diffusivity $D_1$ and a smaller one $D_2$) \cite{MiyaguchiAkimotoYamamoto:2016,UneyamaMiyaguchiAkimoto:2019}. Since the diffusivity $D$ is time-independent, the EAMSD in Eq. \eqref{EAMSD} presents a normal diffusion
\begin{equation}\label{EAMSD1}
  \langle x^2(t)\rangle = 2\langle D\rangle t.
\end{equation}
For TAMSD, the integral in Eq. \eqref{TAMSD} can be understood in a discrete form with a sum of diffusivity at time series $t_i$, i.e.,
\begin{equation}\label{TAMSD-Disc1}
  \overline{\delta^2(\Delta)}\simeq \frac{2\Delta}{N}\sum_{i=1}^N D_i,
\end{equation}
with all $D_i$ are independent and have the identical distribution as random variable $D$. The number of $D_i$, $N$, tends to infinity as measurement $T\rightarrow\infty$, and then the sum of $D_i$ divided by $N$ converges to the ensemble average $\langle D\rangle$, i.e.,
\begin{equation}\label{TAMSD1}
  \overline{\delta^2(\Delta)}\simeq 2\langle D\rangle\Delta.
\end{equation}
This result is consistent to the EAMSD in Eq. \eqref{EAMSD1}, which implies the ergodic nature of the superstatistical model. Both the vast trajectories in EAMSD and the very long measurement time of individual trajectory have the averaging effect on the random diffusivity $D$, and make the Langevin system \eqref{DDmodel} converge to the classical Brownian motion with a constant diffusivity $\langle D\rangle$.

Furthermore, since TAMSD converges to its mean in Eq. \eqref{TAMSD1}, the corresponding EB parameter must tend to zero as $T\rightarrow\infty$. More specifically, combining the definition of the EB parameter in Eq. \eqref{Def-EB} and the TAMSD in Eq. \eqref{TAMSD-Disc1}, we obtain
\begin{equation}\label{EB-case1}
  \textrm{EB}\simeq \frac{1}{N}\left[\frac{\langle D^2\rangle}{\langle D\rangle^2}-1\right].
\end{equation}
See the details in Appendix \ref{App0}.
Since the number $N$ is proportional to the measurement time $T$, the EB parameter in Eq. \eqref{EB-case1} decays with rate $T^{-1}$.

\begin{figure*}
  \centering
  \includegraphics[scale=0.375]{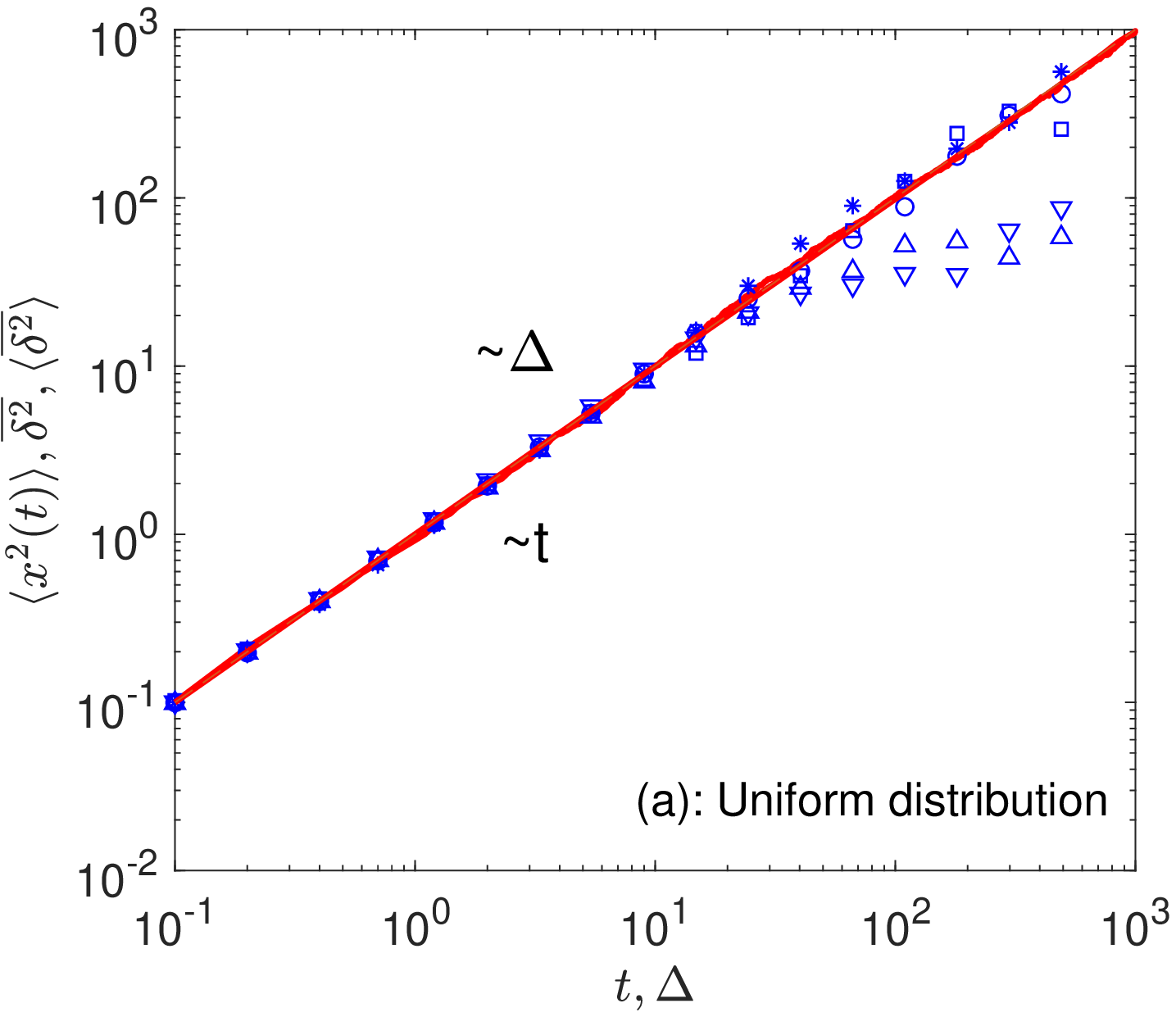}
  \includegraphics[scale=0.375]{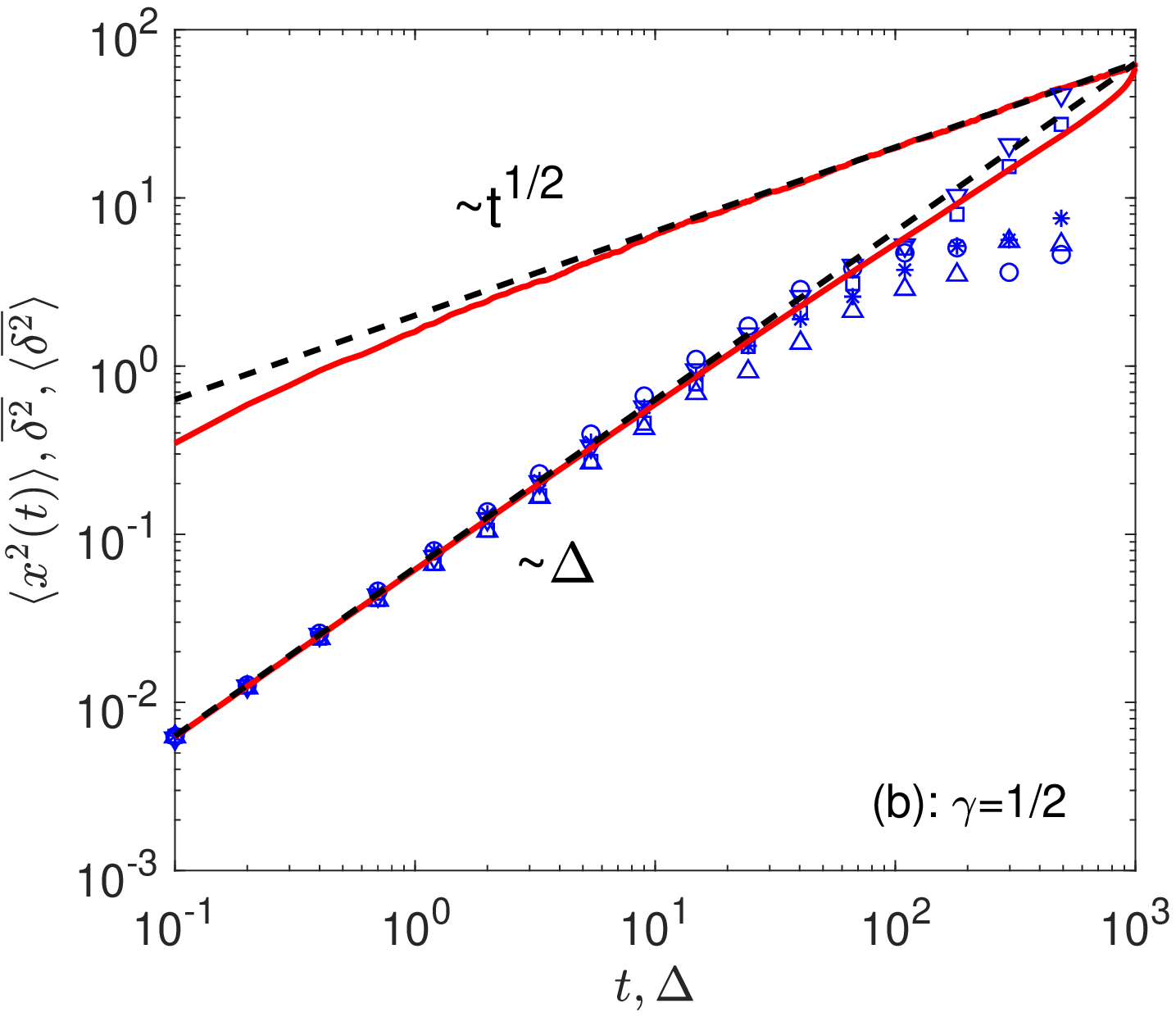}
  \includegraphics[scale=0.375]{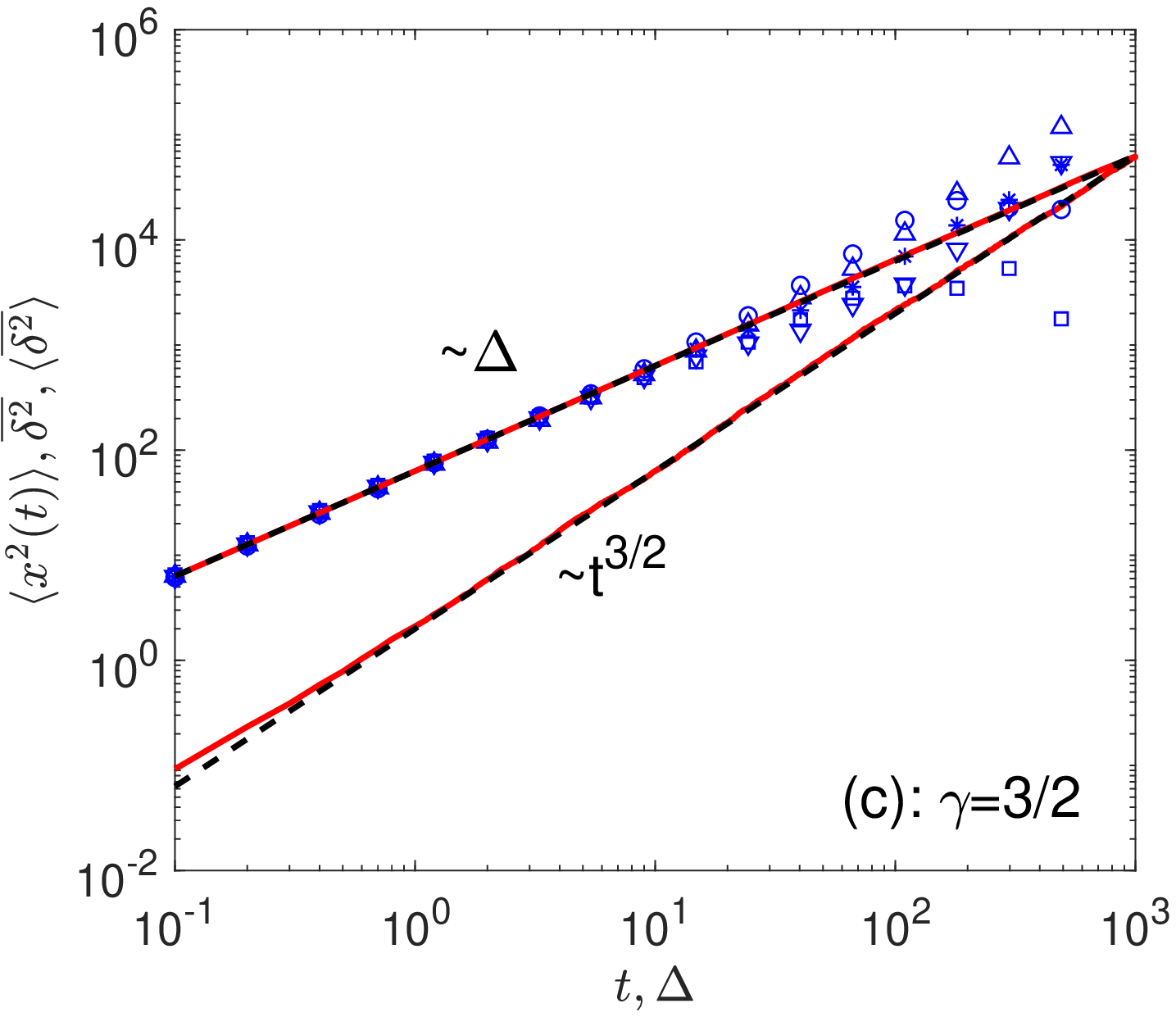}
  \caption{(Color online) EAMSD $\langle x^2(t)\rangle$ and ensemble-averaged TAMSD $\langle\overline{\delta^2(\Delta)}\rangle$ (red solid lines) as well as five individual time traces $\overline{\delta^2(\Delta)}$ (blue markers) for Langevin system with random diffusivity $D(t)$ in Eq. \eqref{DDmodel}. The theoretical results for $\langle x^2(t)\rangle$ in Eq. \eqref{EAMSD} and $\langle\overline{\delta^2(\Delta)}\rangle$ in Eq. \eqref{EATAMSD-Asym} are shown by black dashed lines. The theoretical results often superimpose with the results of simulations. Especially, four lines coincide in (a). The diffusivity is a random variable obeying uniform distribution on the interval $[0,1]$ in (a), while $D(t)$ obeys the exponential distribution with mean $\gamma t^{\gamma-1}$ in (b) for $\gamma=1/2$ and in (c) for $\gamma=3/2$. Other parameters: the measurement time is $T=10^3$, the number of trajectories used for ensemble is $10^3$, and the initial position is $x_0=0$.
  }\label{fig1}
\end{figure*}

In simulations for the superstatistical scenario, we take the example of $D$ obeying uniform distribution on the interval $[0,1]$. The same normal diffusion for EAMSD $\langle x^2(t)\rangle$ and ensemble-averaged TAMSD $\langle\overline{\delta^2(\Delta)}\rangle$ are observed in Fig. \ref{fig1} (a). In addition, we also show five individual time traces in this figure. The coincidence for short $\Delta$ ranging from $1/10$ to $10$ implies the ergodicity of the Langevin system Eq. \eqref{DDmodel}, while the discrepancy between different individual traces for large $\Delta$ is caused by the failure of the condition $\Delta\ll T$.
The EB parameter for this example is shown in Fig. \ref{fig2} with circle markers. The decay rate in simulation is $T^{-1}$, which agrees with the theoretical result in Eq. \eqref{EB-case1}.

\subsection{Uncorrelated diffusivity}
The second major category of diffusivities $D(t)$ are both random and time-dependent, but uncorrelated between different times. We assume the first moment of diffusivity is
\begin{equation}\label{Mean-Dt}
  \langle D(t)\rangle= \gamma t^{\gamma-1}(\gamma\neq1),
\end{equation}
which is equal to $f(t)$ in the exponential distribution in Eq. \eqref{time-dependet-dist}. The second moment of $D(t)$ is assumed to be finite for a given time $t$, which is valid for exponential distribution, or other exponential-type distributions, such as the Rayleigh distribution and the stretched Gaussian distribution. Then, the EAMSD in Eq. \eqref{EAMSD} is anomalous:
\begin{equation}
  \langle x^2(t)\rangle \simeq 2t^\gamma.
\end{equation}

Since the diffusivity $D(t)$ is uncorrelated between different times, similar to Eq. \eqref{TAMSD-Disc1}, the integral in Eq. \eqref{TAMSD} can also be understood in a discrete form with a sum of diffusivity at time series $t_i$, i.e.,
\begin{equation}\label{TAMSD-Disc2}
  \overline{\delta^2(\Delta)}\simeq \frac{2\Delta}{N}\sum_{i=1}^N D(t_i)
\end{equation}
with all $D(t_i)$ being random variables but uncorrelated.
This time-dependent scenario is more difficult than the case of Eq. \eqref{TAMSD-Disc1}, since $D(t_i)$ have different mean and variance.
Therefore, the distribution of $\overline{\delta^2(\Delta)}$ with time-dependent diffusivity seems difficult to be obtained. Fortunately, by use of the Law of Large Numbers \cite{Feller:1971}, if the variance of all $D(t_i)$ are finite for any time $t_i$, then the average of diffusivity $D(t_i)$ converges to its ensemble average, i.e., $\overline{D(t)} = \langle \overline{D(t)} \rangle$ as $T\rightarrow\infty$.

Here, we will loosen the condition of the Law of Large Numbers since the variance of $D(t_i)$ might tend to infinity (e.g., the exponential distribution in Eq. \eqref{time-dependet-dist}). Our assumption is that the variance of $D(t_i)$ does not grow too fast, just in a speed slower than the square of mean multiplied by a linearly growing rate $t$, i.e.,
\begin{equation}
  \frac{\langle D^2(t)\rangle}{\langle D(t)\rangle^2} \ll t
\end{equation}
for large $t$.
More specific elaborations will be presented when we calculate the EB parameter in Eq. \eqref{EB-case2}. And now let us focus on the result by using the Law of Large Numbers.

By virtue of the Law of Large Numbers, we know that each individual trajectory of the TAMSD is self-averaged, i.e.,
\begin{equation}\label{Case2-2}
  \overline{\delta^2(\Delta)}\simeq 2\Delta \langle \overline{D(t)} \rangle,
\end{equation}
as $T\rightarrow\infty$.
This is reminiscent of an extension of Brownian motion, scaled Brownian motion, which is anomalous with a nonlinearly growing EAMSD $\langle x^2(t)\rangle\simeq 2t^\gamma$ due to time-dependent but determined diffusivity $D(t)= \gamma t^{\gamma-1}(\gamma\neq1)$. Its TAMSD is still normal
\begin{equation}\label{Case1-2}
  \overline{\delta^2(\Delta)}\simeq 2\Delta T^{\gamma-1}
\end{equation}
as Eqs. \eqref{TAMSD} and \eqref{Case2-2} show. Albeit the TAMSD of scaled Brownian motion appears to be nonrandom in the limit of $\Delta\ll T$, implying a reproductive behavior between individual trajectories, the disparity between time and ensemble averages $\langle x^2(\Delta)\rangle\neq \overline{\delta^2(\Delta)}$ indicates that scaled Brownian motion is a nonergodic process \cite{JeonChechkinMetzler:2014}.

Similar to the superstatistical approach, where the Brownian motion with a random diffusivity converge to the classical Brownian motion in the sense of EAMSD and TAMSD, the random diffusivity model with time-dependent diffusivity converge to scaled Brownian motion in the long time limit.
They are equivalent when the means of $D(t)$ in the two situations, random or deterministic diffusivities, are the same. Their TAMSDs both present the reproductive nature. However, that the Langevin system Eq. \eqref{DDmodel} is ergodic or not is determined by whether the (mean of) diffusivity $D(t)$ is time-dependent.

The result Eq. \eqref{Case2-2} can be justified if the corresponding EB parameter tends to zero as $T\rightarrow\infty$. From the definition of EB parameter in Eq. \eqref{Def-EB}, we find
\begin{equation}\label{EB-case2}
  \textrm{EB}=\frac{\int_0^T\langle D^2(t)\rangle-\langle D(t)\rangle^2 dt}
  {\left[\int_0^T\langle D(t)\rangle dt\right]^2},
\end{equation}
with the details presented in Appendix \ref{App1}. When $D(t)$ becomes time-independent, Eq. \eqref{EB-case2} recovers to Eq. \eqref{EB-case1} in the superstatistical scenario. The long time limit of Eq. \eqref{EB-case2} is
\begin{equation}
  \textrm{EB}\simeq \frac{E^2(T)}{T},
\end{equation}
where
\begin{equation}
  E^2(T)=\frac{\langle D^2(T)\rangle-\langle D(T)\rangle^2}{\langle D(T)\rangle^2}
\end{equation}
only depends on the first two moments of diffusivity $D(T)$ at long time. $E(T)$ is named as coefficient of variation (CV), which is a dimensionless variable and defined as the ratio of standard deviation to the mean. A larger CV usually corresponds to a greater dispersion of a set of data. The CV has been used in many branches of probability theory, such as renewal theory, queuing theory and reliability theory.

Only if the square of CV grows with a rate slower than $T$, then $\textrm{ET}\rightarrow0$ as $T\rightarrow\infty$. This condition can be satisfied by most of kinds random variables which only take positive values. Especially for the ones with single mode, where the second moment behaves as the square of its mean and CV is a $T$-independent constant, such as the exponential distribution, Rayleigh
distribution, Erlang distribution and superexponential distribution, etc. In these cases, the CV is a constant and the EB parameter in Eq. \eqref{EB-case2} decays to zero as
\begin{equation}\label{EBT-1}
  \textrm{EB}\simeq \mathcal{O}(T^{-1}),
\end{equation}
similar to many ergodic processes, including Brownian motion ($H=1/2$), fraction Brownian motion and fractional Langevin equation with Hurst index ($H<3/4$) \cite{DengBarkai:2009}. Even for the processes with multiple modes, usually resulting from power-law distributed waiting time, such as occupation time statistics in ergodic CTRW \cite{SchulzBarkai:2015}, and enhanced L\'{e}vy walk displaying strong anomalous diffusion behavior \cite{RebenshtokDenisovHanggiBarkai:2014,WangChenDeng:2020}, albeit their CVs grow as $T$ increases in a rate depending on the specific power-law exponent, this rate is slower than $T^{1/2}$ and EB parameter vanishes for long time limit.

\begin{figure}
  \centering
  \includegraphics[scale=0.5]{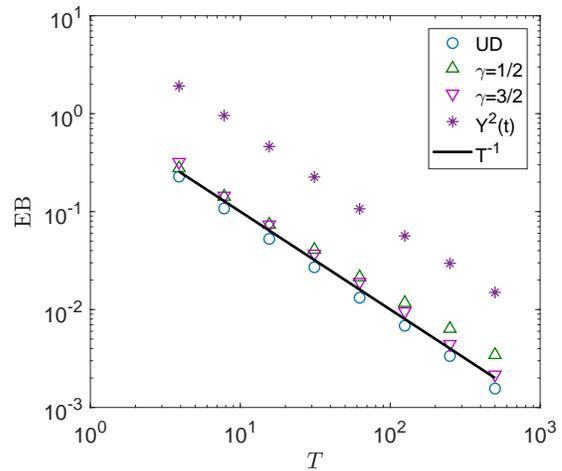}\\
  \caption{(Color online) EB parameters for four kinds of diffusivity $D(t)$: circle markers for the uniform distribution on the interval $[0,1]$, triangle markers for the exponential distribution with mean $\gamma t^{\gamma-1}$, and star markers for the square of OU process ($D(t)=Y^2(t)$). The markers are all decaying as the reciprocal of measurement $T$, parallel to the auxiliary line $T^{-1}$. Other parameters: the lag time is $\Delta=1$, the number of trajectories used for ensemble is $10^3$, and the initial position is $x_0=0$.
  }\label{fig2}
\end{figure}

We make some simulations on MSDs in this part by assuming the uncorrelated diffusivity $D(t)$ obeying exponential distribution in Eq. \eqref{time-dependet-dist} with the mean $\gamma t^{\gamma-1}$ ($\gamma=1/2$ in Fig. \ref{fig1} (b) and $\gamma=3/2$ in Fig. \ref{fig1} (c)). The EAMSD exhibits subdiffusion for $\gamma<1$ in (b) and superdiffusion for $\gamma>1$ in (c), while the ensemble-averaged TAMSD presents normal diffusion for any $\gamma$. This discrepancy implies the nonergodic behavior of Langevin system Eq. \eqref{DDmodel} when $\gamma\neq1$.
In addition, similar to the superstatistical scenario in Fig. \ref{fig1} (a), five individual time traces are shown in both (b) and (c). The coincidence between different individual traces can be observed when $\Delta\ll T$.
The EB parameter for this example is shown in Fig. \ref{fig2} with triangle markers (positive triangle for $\gamma=1/2$ and inverted triangle for $\gamma=3/2$). The decay rates in simulations are both $T^{-1}$, which agrees with the theoretical result in Eq. \eqref{EBT-1}.

\subsection{Correlated diffusivity}
The last major category is that diffusivity $D(t)$ is a stochastic process, which is autocorrelated between different times. Similar to the previous cases, the EAMSD and ensemble-averaged TAMSD only depend on the mean of diffusivity $\langle D(t)\rangle$, i.e., the Eqs. \eqref{EAMSD} and \eqref{EATAMSD-Asym} are still valid here. However, the TAMSD is not reproductive and $\phi(\eta)\neq\delta(\eta-1)$ any more. The distribution $\phi(\eta)$ will present different shapes for various stochastic process $D(t)$. As Eq. \eqref{TAMSD} shows, the TAMSD for sufficiently large $T$ does make an average on the white noise $\xi(t)$ which has $\delta$-correlated function. In contrast, the correlated diffusivity $D(t)$ remains in the expression of TAMSD in Eq. \eqref{TAMSD}. In this sense, the distribution of TAMSD is mainly determined by the correlation of displacements at different time.

One typical example is that $D(t)$ is a renewal process, which defines the total number of renewal events happening before time $t$. Its mathematical definition is
\begin{equation}\label{D-delta}
  D(t)=\sum_{i=1}^\infty \delta(t-t_i),
\end{equation}
where $t_i$ are random variables describing the times at which the $i$-th renewal event occurs.
The time difference between two successive renewal events can be regarded as the waiting time in the framework of CTRW \cite{LubelskiSokolovKlafter:2008,HeBurovMetzlerBarkai:2008,BurovMetzlerBarkai:2010}. The power-law distributed waiting times characterize the long trapping events and thus lead to the subdiffusive CTRW. The corresponding limiting distribution $\phi(\eta)$ of renewal times as measurement time $T\rightarrow\infty$ has been discussed in many references \cite{HeBurovMetzlerBarkai:2008,AghionKesslerBarkai:2019,LeibovichBarkai:2019,WangDengChen:2019}, where the dimensionless random variable $\eta$ obeys the Mittag-Leffler distribution, i.e.,
\begin{equation}\label{MLD}
  \lim_{T\rightarrow\infty}\phi(\eta)= \frac{\Gamma^{1/\alpha(1+\alpha)}}{\alpha\eta^{1+1/\alpha}}L_\alpha
  \left[\frac{\Gamma^{1/\alpha}(1+\alpha)}{\eta^{1/\alpha}}\right].
\end{equation}
Here, $\alpha$ is the exponent of power-law distribution, and $L_\alpha(1)$ is the one-sided L\'{e}vy stable distribution whose Laplace pair is $\exp(-u^\alpha)$ \cite{Feller:1971,Barkai:2001}.
For the $\phi(\eta)$ in Eq. \eqref{MLD}, the corresponding EB parameter is \cite{HeBurovMetzlerBarkai:2008}
\begin{equation}
  \textrm{EB} = \frac{2\Gamma^2(1+\alpha)}{\Gamma(1+2\alpha)}-1.
\end{equation}

For other kinds of diffusivity $D(t)$, the explicit form of PDF $\phi(\eta)$ and EB parameter cannot be easily obtained. Instead, the direct numerical simulations on the diffusing diffusivity $D(t)$ can be regarded as a theoretical reference object to justify Eq. \eqref{TAMSD}. Here, we mainly take the three common examples mentioned in Sec. \ref{Sec2}, i.e., the square of Brownian motion $D(t)=B^2(t)$, the reflected Brownian motion $D(t)=|B(t)|$, and the square of OU process $D(t)=Y^2(t)$.

We show the explicit results of some important quantities in Table \ref{table1}, including the means of diffusivity $\langle D(t)\rangle$, the corresponding EAMSDs and ensemble-averaged TAMSDs, and the asymptotic behaviors of EB parameter of the three kinds of diffusing diffusivity $D(t)$. Among them, the MSDs can be directly evaluated by use of Eqs. \eqref{EAMSD} and \eqref{EATAMSD-Asym}. It can be seen that the first two cases are nonergodic while $D(t)=Y^2(t)$ is ergodic. The ensemble-averaged TAMSDs are all exhibiting normal diffusion ($\Delta$). By contrast, the EAMSD can be normal or anomalous, which depends on whether $\langle D(t)\rangle$ converges to a constant for long time.

The EB parameter in the case of $D(t)=B^2(t)$ satisfies
\begin{equation}\label{EB-case3}
  \lim_{T\rightarrow\infty}\textrm{EB}=\frac{4}{3}.
\end{equation}
The detailed calculations are presented in Appendix \ref{App2}. But the EB parameter for $D(t)=|B(t)|$ cannot be obtained easily. We fit the simulation data and find an approximation result $\textrm{EB}\rightarrow13/40$ as $T\rightarrow\infty$. In contrast to the above two cases, the case with $D(t)=Y^2(t)$ presents us a quite different phenomenon, where EB parameter tends to zero. The inherent mechanism attributes to the ergodicity of OU process since it reaches a stationary steady for long time \cite{WangDengChen:2019}. Diffusivity can be regarded as an observable of OU process, and its time average will converge to the corresponding ensemble average \cite{BurovMetzlerBarkai:2010}. Therefore, this case recovers the results of $\textrm{EB}\rightarrow0$ as the uncorrelated diffusivities present in the previous section. In addition, since the OU process only has single mode for long time, the EB parameter decays with rate $T^{-1}$ as Eq. \eqref{EBT-1} shows.

\begin{table}
\caption{Mean of diffusivity $D(t)$, EAMSD, ensemble-averaged TAMSD, and EB parameter at $T\rightarrow\infty$ for three kinds of correlated $D(t)$.}\label{table1}
\scalebox{1.20}{
\begin{tabular}{ccccc}
  \hline
  $D(t)$ & $\langle D(t)\rangle$ & $\langle x^2(t)\rangle$ & $\langle\overline{\delta^2(\Delta)}\rangle$ & EB \\
\hline
  $B^2(t)$ & $t$ & $t^2$ & $\Delta T$ & $=\frac{4}{3}$ \\

  $|B(t)|$ & $\sqrt{\frac{2}{\pi}}t^{1/2}$ & $\sqrt{\frac{32}{9\pi}}t^{3/2}$ & $\sqrt{\frac{32}{9\pi}}\Delta T^{1/2}$ & $\approx\frac{13}{40}$ \\

  $Y^2(t)$ & $(1-e^{-2t})/2$ & $t$ & $\Delta$ & $\propto T^{-1}$\\
  \hline
\end{tabular}}
\end{table}

\begin{figure*}
  \centering
  \includegraphics[scale=0.375]{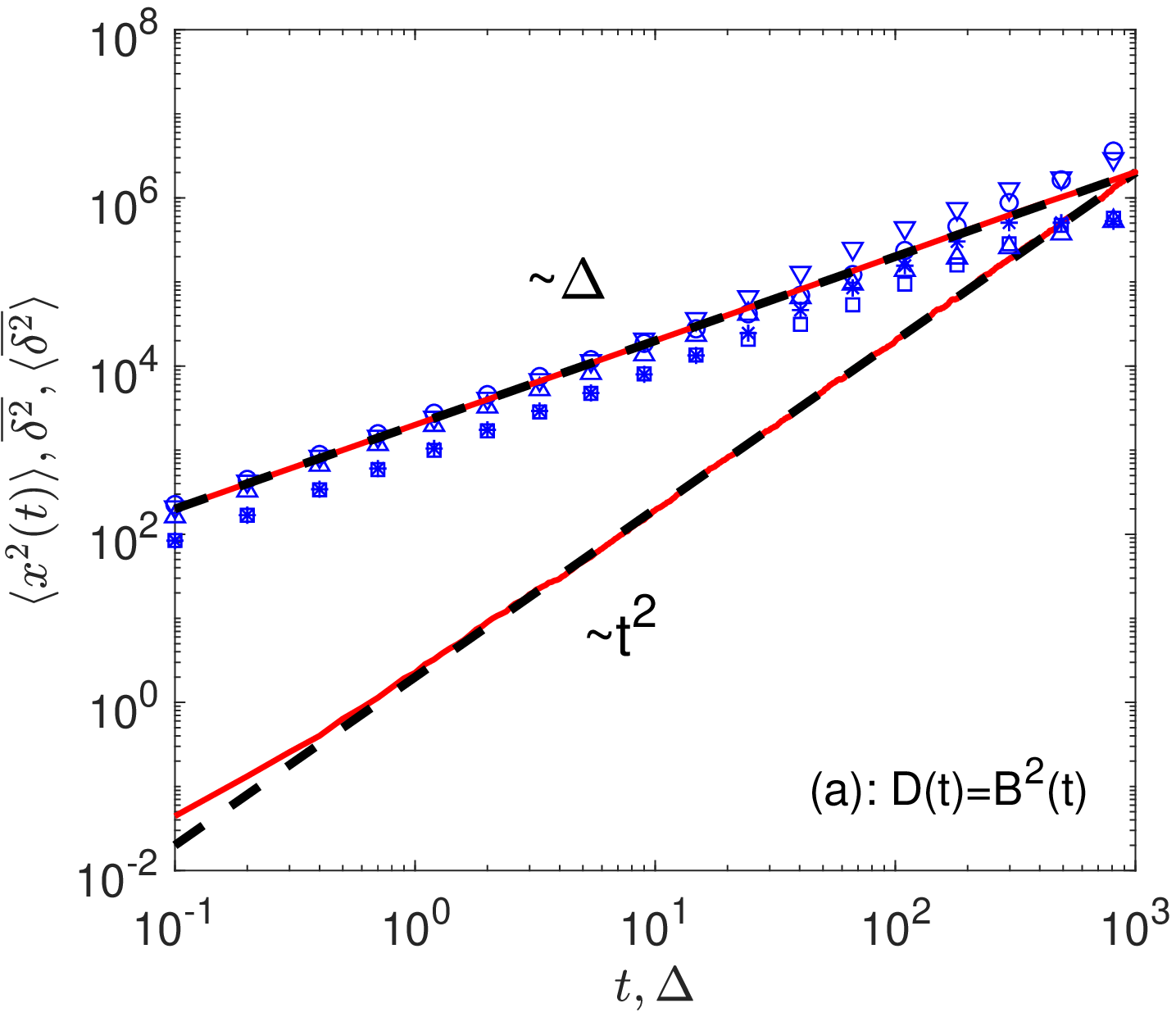}
  \includegraphics[scale=0.375]{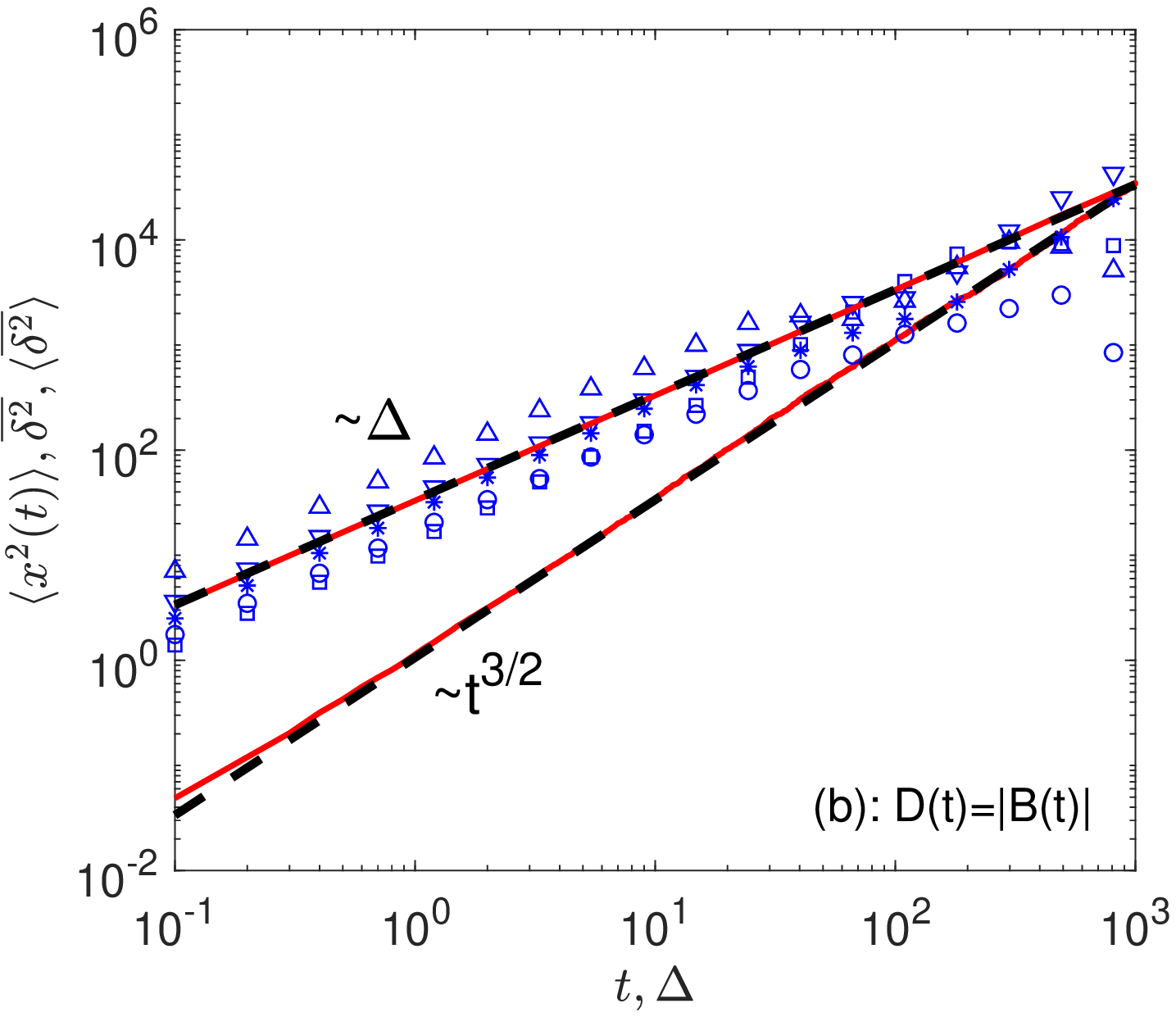}
  \includegraphics[scale=0.375]{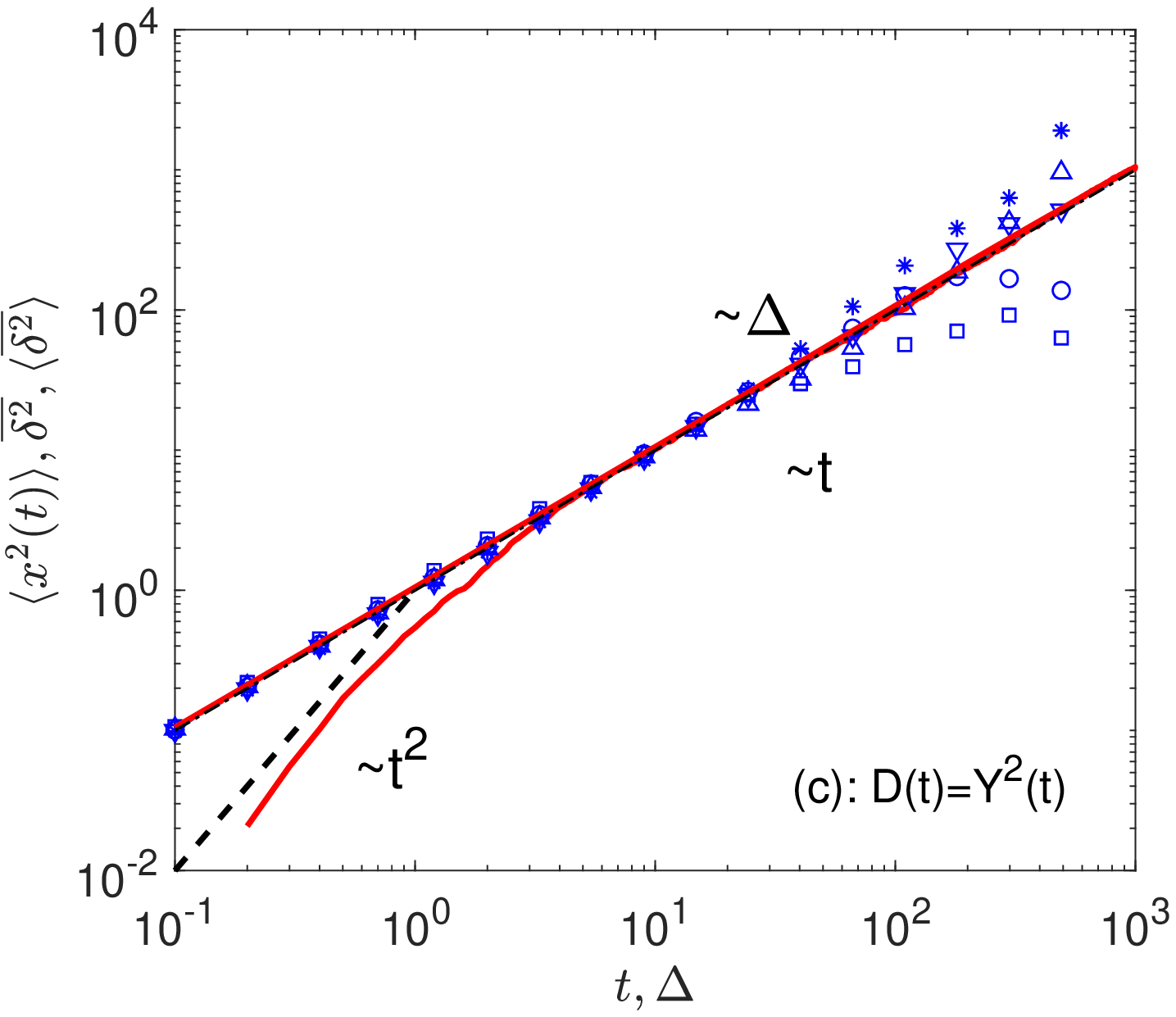}
  \caption{(Color online) EAMSD $\langle x^2(t)\rangle$ and ensemble-averaged TAMSD $\langle\overline{\delta^2(\Delta)}\rangle$ (red solid lines) as well as five individual time traces $\overline{\delta^2(\Delta)}$ (blue markers) for Langevin system with random diffusivity $D(t)$ in Eq. \eqref{DDmodel}. The theoretical results for $\langle x^2(t)\rangle$ in Eq. \eqref{EAMSD} and $\langle\overline{\delta^2(\Delta)}\rangle$ in Eq. \eqref{EATAMSD-Asym} are shown by black dashed lines. The theoretical results often superimpose with the results of simulations. Especially, four lines coincide for long time and an auxiliary line $t^2$ is added to indicate the asmptotics of EAMSD for short time in (c). The diffusing diffusivity $D(t)$ is taken as $B^2(t)$ in (a), $|B(t)|$ in (b) and $Y^2(t)$ in (c), where $B(t)$ is Brownian motion and $Y(t)$ is OU process. Other parameters: the measurement time is $T=10^3$, the number of trajectories used for ensemble is $10^3$, and the initial position is $x_0=0$.}\label{fig3}
\end{figure*}

\begin{figure*}
  \centering
  \includegraphics[scale=0.375]{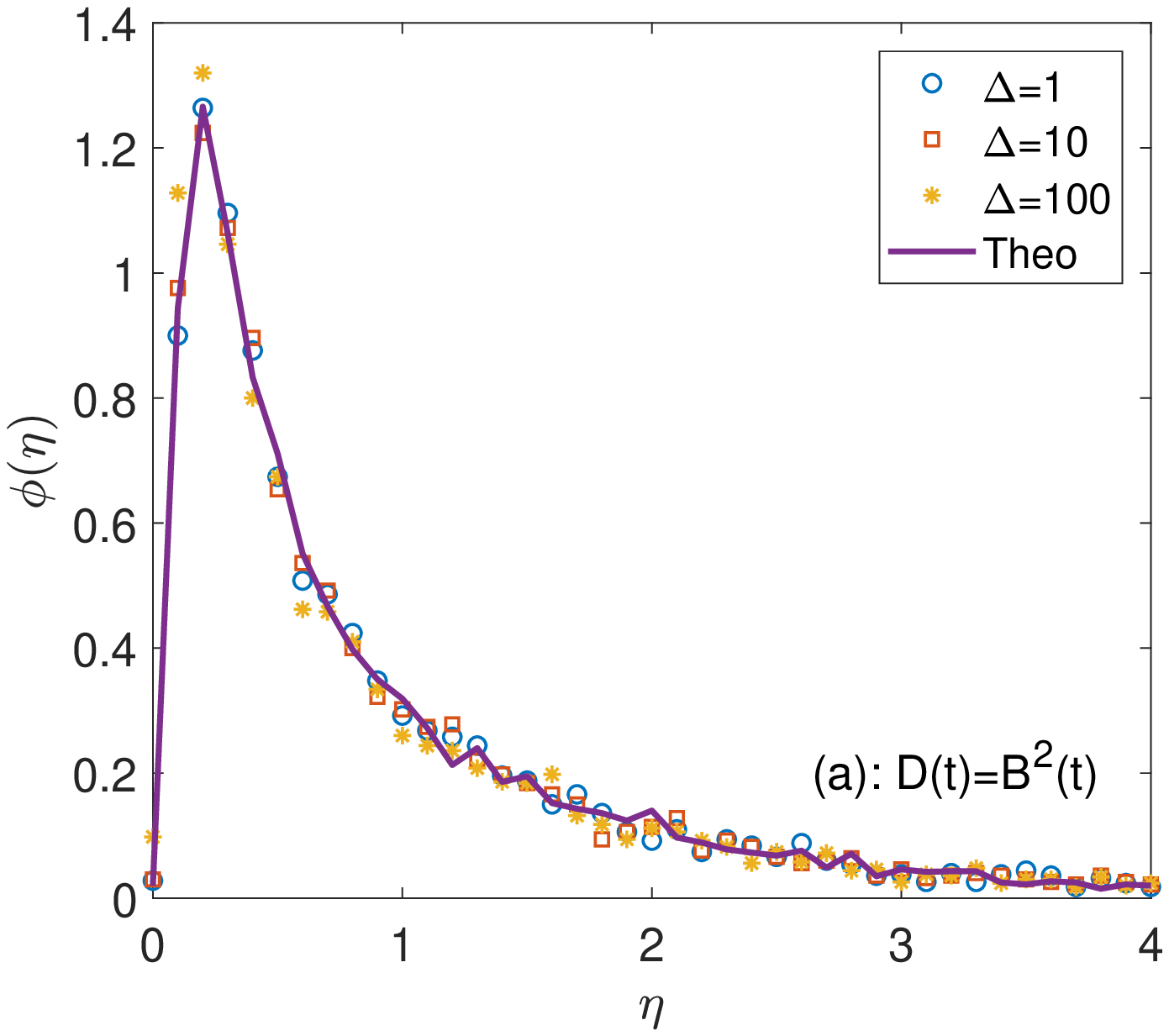}
  \includegraphics[scale=0.375]{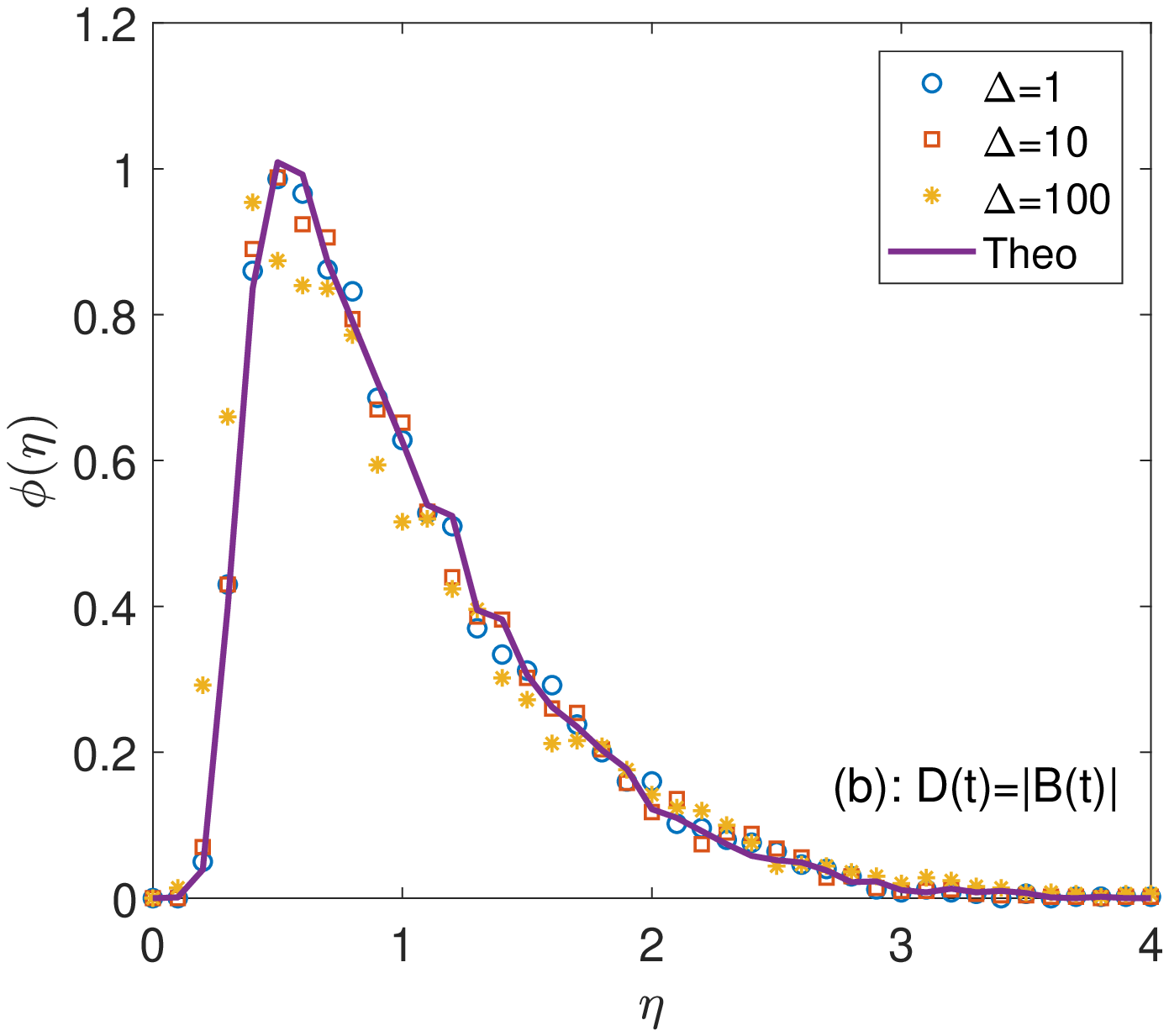}
  \includegraphics[scale=0.375]{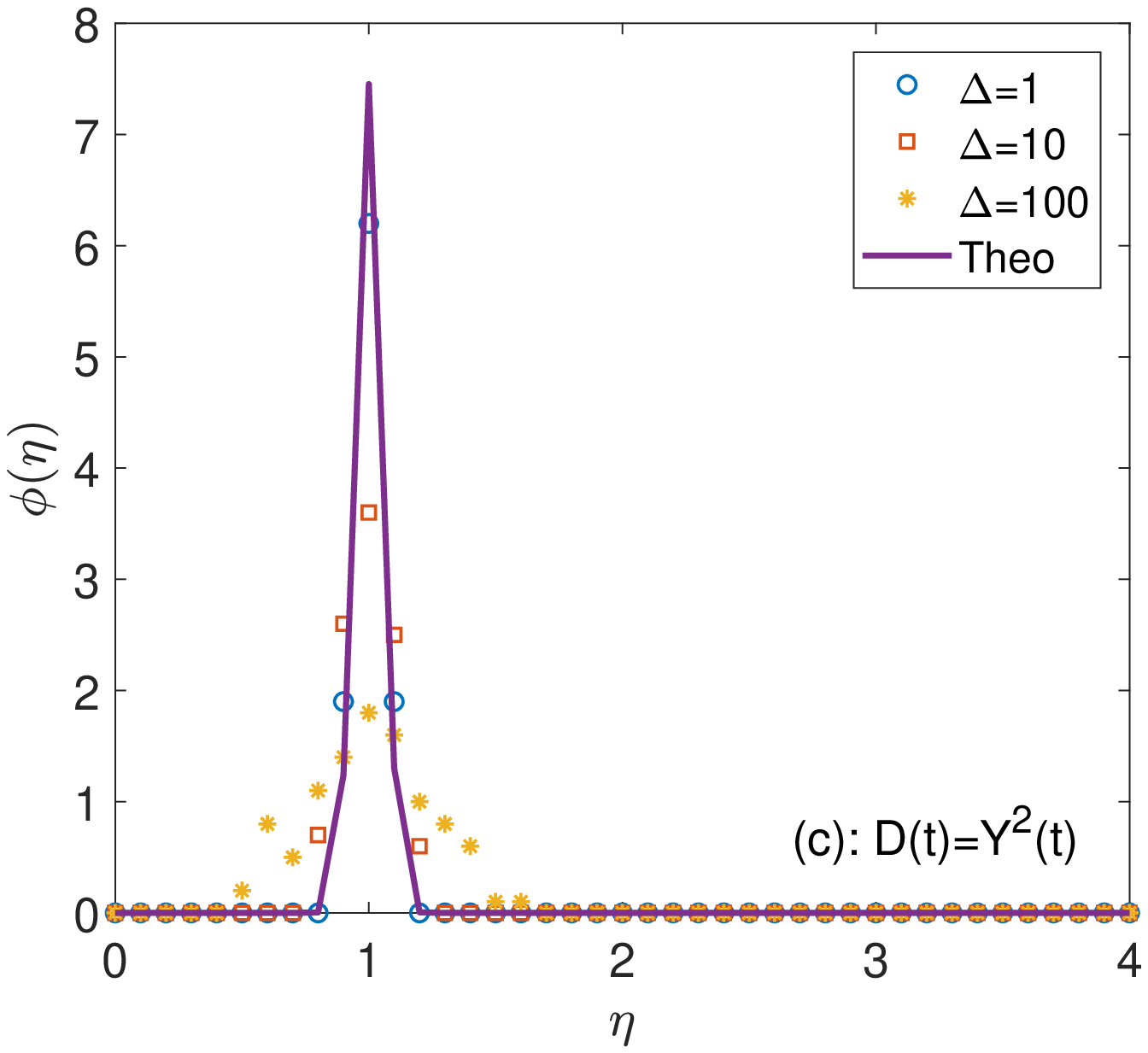}
  \caption{(Color online) Amplitude scatter PDF $\phi(\eta)$ for three kind of diffusing diffusivity: $D(t)=B^2(t)$ in (a), $D(t)=|B(t)|$ in (b), and $D(t)=Y^2(t)$ in (c), where $B(t)$ is Brownian motion and $Y(t)$ is OU process. The markers (circle, square, star) denote the simulations for $\Delta$ ($=1,10,100$), respectively. The solid lines are obtained by making simulations directly on the trajectories of diffusivity $D(t)$ based on the theoretical result $\eta
  \simeq \overline{D(t)}/\langle \overline{D(t)}\rangle$ in Eq. \eqref{Def-eta}. Due to the condition $\Delta\ll T$, the circle markers ($\Delta=1$) are more consistent than the star markers ($\Delta=100$) with the solid lines. The dimensionless variable $\eta$ has a broad distribution in (a) and (b), but a narrow distribution (similar to $\delta$-function) in (c). Other parameters: the measurement time is $T=10^4$, the number of trajectories used for ensemble is $10^4$, and the initial position is $x_0=0$.
  }\label{fig4}
\end{figure*}

\begin{figure}
  \centering
  \includegraphics[scale=0.5]{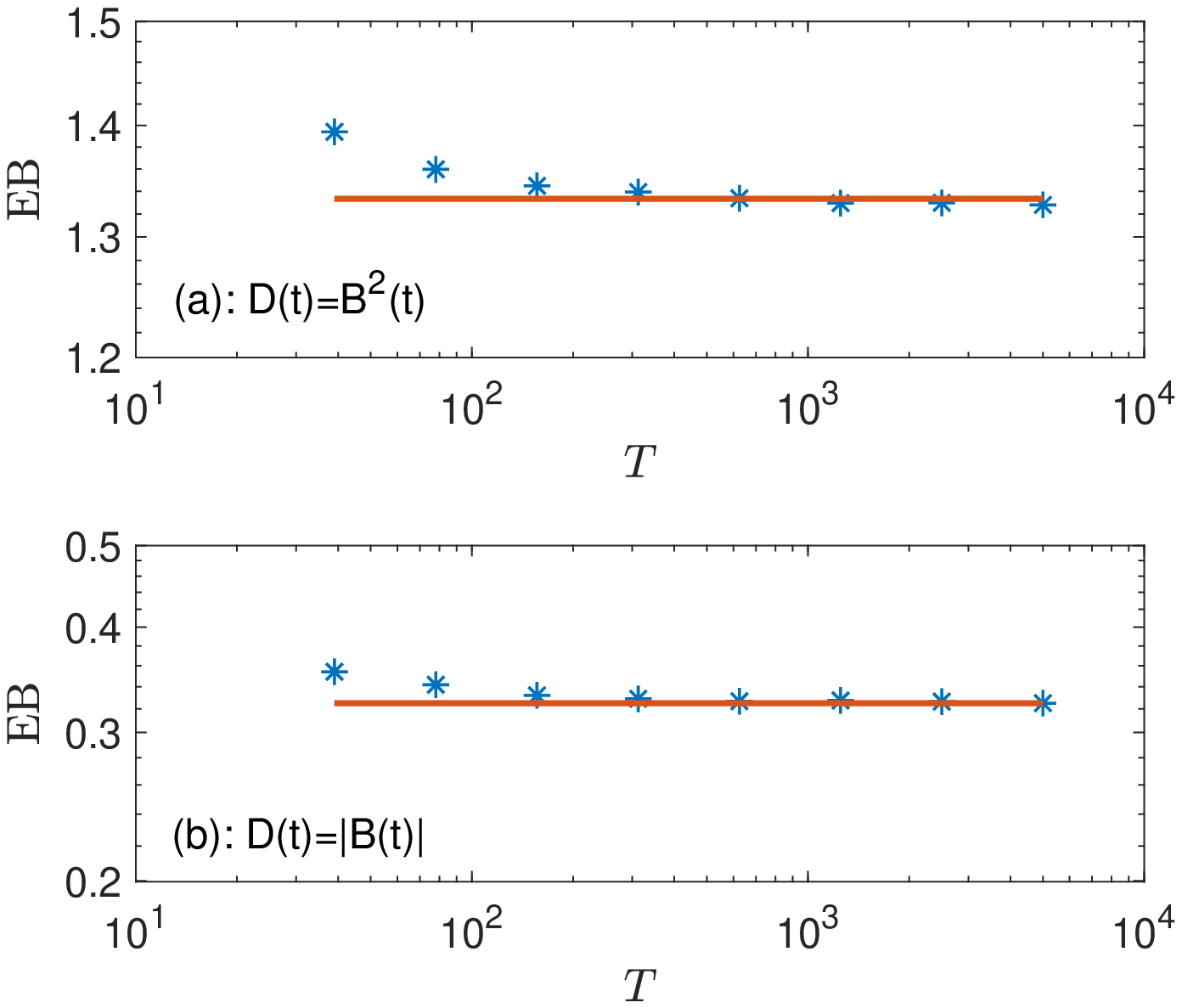}
  \caption{(Color online) EB parameters for two kinds of diffusing diffusivity: $D(t)=B^2(t)$ in (a) and $D(t)=|B(t)|$ in (b). Instead of decaying to zero, the star markers converge to a constant for long time. This constant agrees with the theoretical result in Eq. \eqref{EB-case3} in (a) and approximates $13/40$ in (b).
  Other parameters: the lag time is $\Delta=1$, the number of trajectories used for ensemble is $10^4$, and the initial position is $x_0=0$.
  }\label{fig5}
\end{figure}

In Figs. \ref{fig3}, \ref{fig4} and \ref{fig5}, respectively, we show the MSDs, the amplitude scatter PDF $\phi(\eta)$ and the corresponding EB parameters for Langevin system with correlated diffusivities: $D(t)=B^2(t)$ in (a), $D(t)=|B(t)|$ in (b) and $D(t)=Y^2(t)$ in (c). In Fig. \ref{fig3}, the simulations of EAMSD and ensemble-averaged TAMSD agree with the corresponding theoretical results in Eqs. \eqref{EAMSD} and \eqref{EATAMSD-Asym} completely. For five individual time traces $\overline{\delta^2(\Delta)}$, the case $D(t)=Y^2(t)$ in (c) presents a similar phenomenon to Fig. \ref{fig1}, i.e., they coincide with each other when $\Delta\ll T$, which implies the ergodic behavior as $T\rightarrow\infty$. In addition, this ergodic nature for $D(t)=Y^2(t)$ can be also indicated by the narrow distribution (similar to $\delta(\eta-1)$) in Fig. \ref{fig4} (c) and the decaying EB parameter in the speed of $T^{-1}$ in Fig. \ref{fig2} (the star markers). The reason of the narrow distribution (rather than a $\delta$-function) is that the measurement time $T$ is finite in simulations. It cannot really go to the infinity in simulations or experiments. More discussions on the effect of the finite measurement time $T$ can be found in Refs. \cite{GodecMetzler:2013,JeonMetzler:2012,JeonMetzler:2010-2}.

In contrast to $D(t)=Y^2(t)$ in (c), the cases $D(t)=B^2(t)$ in (a) and $D(t)=|B(t)|$ in (b) present the obvious nonergodic behaviors. In Fig. \ref{fig3}, the EAMSDs exhibit superdiffusion ($t^2$ in (a) and $t^{3/2}$ in (b)), whereas the ensemble-averaged TAMSDs are both normal. In addition, the individual time traces (blue markers) do not coincide any more even for $\Delta\ll T$.

In Fig. \ref{fig4}, the broad distributions of $\phi(\eta)$ in (a) and (b) also indicate the nonreproductive behavior between
individual trajectories. The theoretical results (solid lines) are obtained by making simulations directly on the trajectories of diffusing diffusivity $D(t)$ without simulating the Langevin equation Eq. \eqref{DDmodel}. Then based on the result $\eta
  \simeq \overline{D(t)}/\langle \overline{D(t)}\rangle$ in Eq. \eqref{Def-eta}, we can draw a graph about the distribution $\phi(\eta)$ by use of the trajectories of $D(t)$.
The $\phi(\eta)$ decays to zero at $\eta=0$ in (a) and (b), different to the nonzero value of subdiffusive CTRW resulting from completely immobilized particles \cite{HeBurovMetzlerBarkai:2008,LubelskiSokolovKlafter:2008,BurovJeonMetzlerBarkai:2011}. The overall shapes of the $\phi(\eta)$ here are
similar to the ones of superdiffusive heterogeneous diffusion processes in Ref. \cite{CherstvyChechkinMetzler:2013}, where the diffusivity is a determined space-dependent function $D(x)$ and the parameters in TAMSD are fitted with the three-parameter gamma distribution. The possible reasons are the common superdiffusion behavior and heterogeneous Langevin system.

In Fig. \ref{fig5}, the EB parameters for $D(t)=B^2(t)$ and $D(t)=|B(t)|$ converge to a constant ($4/3$ in (a) and approximating $13/40$ in (b)) as $T\rightarrow\infty$, discrepant from the decaying phenomenon in Fig. \ref{fig2}. The nonzero EB parameter also indicates the nonergodic behavior for these two cases.

\section{Summary}\label{Sec5}
\begin{table*}
\caption{Random diffusivity models with different kinds of diffusivity and their (non)ergodic behaviors. For convenience, we assume the mean of diffusivity is $\langle D(t)\rangle\simeq \gamma t^{\gamma-1}$ uniformly for different cases, in which $\gamma=1$ can be regarded as the case of diffusivity being a random variable. We classify the different cases through EAMSD $\langle x^2(t)\rangle$, ensemble-averaged TAMSD $\langle\overline{\delta^2(\Delta)}\rangle$, WEB which means weak ergodicity breaking (Yes or No), the scatter of dimensionless amplitude $\phi(\eta)$ and EB parameter when $T\rightarrow\infty$. \\}\label{table2}
\resizebox{13cm}{!}{
\begin{tabular}{cccccc}
  \hline
  Different cases & ~~$\langle x^2(t)\rangle$ & ~~$\langle\overline{\delta^2(\Delta)}\rangle$ & ~~WEB & ~~$\phi(\eta)$ & ~~EB($T\rightarrow\infty$) \\
\hline
  Random variable ($\gamma=1$) & $2t$ & $2\Delta $ & No &  $\delta(\eta-1)$ & $0$ \\

  Uncorrelated diffusivity & $2t^{\gamma}$ & $2\Delta T^{\gamma-1}$ & Yes &  $\delta(\eta-1)$ & $0$  \\

  Correlated diffusivity & $2t^{\gamma}$ & $2\Delta T^{\gamma-1}$ & Yes &  Broad & Constant \\
  \hline
\end{tabular}}
\end{table*}

Brownian yet non-Gaussian phenomena have recently been observed in a large range of complex systems. In contrast to the crossover of PDF from exponential distribution to Gaussian distribution for this new class of diffusion processes, we pay more attention to the ergodic property, especially the scatter of dimensionless amplitude of TAMSD in this paper.

For the sake of detailed discussions on the ergodic property, we consider the random diffusivity model Eq. \eqref{DDmodel} with three categories of diffusivities $D(t)$: a random variable $D$, a time-dependent but uncorrelated diffusivity $D(t)$, and an autocorrelated stochastic process $D(t)$. We find that both EAMSD and ensemble-averaged TAMSD only depend on the mean of diffusivity $\langle D(t)\rangle$ for all kinds of $D(t)$. Different forms of diffusivity $D(t)$ mainly change the property of TAMSD, e.g., the scatter of dimensionless amplitude $\phi(\eta)$ and EB parameter. Note that Eq. \eqref{Def-eta} for the dimensionless variable is valid for any kind of diffusivities. Thus, $\phi(\eta)$ can be directly obtained by use of the information of $D(t)$.
The main results of this paper are presented in Table \ref{table2}, where for convenience we assume $\langle D(t)\rangle\simeq\gamma t^{\gamma-1}$ for all cases. The first scenario that diffusivity is a random variable $D$ can be regarded as the special case of $\gamma=1$.

In Table \ref{table2}, the EAMSD $\langle x^2(t)\rangle$ and ensemble-averaged TAMSD $\langle\overline{\delta^2(\Delta)}\rangle$ are obtained from Eqs. \eqref{EAMSD} and \eqref{EATAMSD-Asym}. The first case is ergodic while another two cases are weakly nonergodic. There is a one-to-one relationship between the scatter of dimensionless amplitude $\phi(\eta)$ and EB parameter. The former converging to $\delta(\eta-1)$ corresponds to the latter converging to zero as $T\rightarrow\infty$, vice versa. In spite the model with uncorrelated diffusivity (the second case) shows a reproductive behavior between individual trajectories since $\textrm{EB}\rightarrow0$, the disparity between anomalous diffusion of EAMSD and normal diffusion of ensemble-averaged TAMSD indicates the weakly nonergodic behavior. For the particular case where the diffusing diffusivity is the square of OU process (i.e., $D(t)=Y^2(t)$), albeit it is a process seemingly belonging to the third case, the truth that it can reach a stationary state makes it possess the properties of the first case (random variable $D$). Therefore, as a result, this kind of process can be assigned to the first group in Table \ref{table2}.

Based on the discussions on the ergodic properties of the random diffusivity model Eq. \eqref{DDmodel}, the potential value of this paper also lies in a possible new characterization of this model to other anomalous diffusion processes that have been discussed or have not yet been discovered. Here are three examples.
(i): As the second case (a time-dependent but uncorrelated diffusivity $D(t)$) says in Sec. \ref{Sec4}, the random diffusivity model converges to scaled Brownian motion for long time. This implies that a large range of processes shares the same statistical phenomena with scaled Brownian motion but their diffusivity might be fluctuating.
(ii): We have shown the equivalence between the subdiffusive CTRW and our model in Eq. \eqref{D-delta} in the third case (a correlation stochastic process $D(t)$). With this in mind, numerous Langevin systems coupled with an (inverse) subordinator can be transferred to the framework of random diffusivity model for the sake of discussion.
(iii): When choosing a diffusing diffusivity, (e.g., $D(t)=B^2(t)$ or $D(t)=|B(t)|$), we find the great similarity between our model and the heterogenous diffusion processes discussed in Ref. \cite{CherstvyChechkinMetzler:2013}. The similarity is embodied in many observables, including the EAMSD (being anomalous), the ensemble-averaged TAMSD (being normal), and the shape of the scatter of dimensionless amplitude $\phi(\eta)$. Note that those two models both describe the motion in an inhomogeneous environment, this seemingly reveals the potential relations between them.
More detailed connections between the random diffusivity model and other common processes will be discussed in the future.

\section*{Acknowledgments}
This work was supported by the Fundamental Research Funds for the Central Universities under grants no. lzujbky-2020-it02, lzujbky-2019-it17.

\appendix
\section{EB parameters for random variable $D$ in Eq. \eqref{EB-case1}}\label{App0}
Combining the definition of dimensionless variable $\eta$ in Eq. \eqref{Def-eta} and the TAMSD in Eq. \eqref{TAMSD-Disc1}, it holds that
\begin{equation}
  \eta\simeq
  \frac{\underset{i=1}{\overset{N}{\sum}} D_i}{\underset{i=1}{\overset{N}{\sum}}\langle D_i\rangle}
  =\frac{\underset{i=1}{\overset{N}{\sum}} D_i}{N\langle D\rangle},
\end{equation}
based on which, one arrives at
\begin{equation}
  \begin{split}
    \langle\eta^2\rangle \simeq \frac{N\langle D^2\rangle+N(N-1)\langle D\rangle^2}{N^2\langle D\rangle^2}.
  \end{split}
\end{equation}
Therefore,
\begin{equation}
  \textrm{EB}=\langle\eta^2\rangle-1 \simeq \frac{1}{N}\left[\frac{\langle D^2\rangle}{\langle D\rangle^2}-1\right].
\end{equation}

\section{EB parameters for uncorrelated diffusing diffusivity $D(t)$ in Eq. \eqref{EB-case2}}\label{App1}
The key to calculate EB parameter is to evaluate the two terms, $\langle(\overline{D(t)})^2\rangle$ and $\langle\overline{D(t)}\rangle$, in Eq. \eqref{EB-Dt}. For the former, since
\begin{equation}\label{App-A1}
  \overline{D(t)}=\frac{1}{T}\int_0^T D(t)dt,
\end{equation}
it holds that
\begin{equation}\label{App-A2}
  \overline{D(t)}^2 = \frac{1}{T^2}\int_0^T\int_0^T D(t_1)D(t_2)dt_1dt_2.
\end{equation}
The two terms $D(t_1)$ and $D(t_2)$ in the integrand are uncorrelated for $t_1\neq t_2$ but correlated for $t_1=t_2$. With this in mind, we have the correlated function
\begin{equation}\label{App-A3}
\begin{split}
     \langle D(t_1)D(t_2)\rangle&=\left[\langle D^2(t_1)\rangle-\langle D(t_1)\rangle^2  \right]\delta(t_1-t_2)  \\
     &~~~+\langle D(t_1)\rangle \langle D(t_2)\rangle.
\end{split}
\end{equation}
Thus we have, by performing the ensemble average on Eq. \eqref{App-A2},
\begin{equation}\label{App-A4}
\begin{split}
    \langle(\overline{D(t)})^2\rangle
    = \frac{1}{T^2}\int_0^T \langle D^2(t)\rangle-\langle D(t)\rangle^2 dt
    + \langle\overline{D(t)}\rangle^2,
\end{split}
\end{equation}
where the last term is
\begin{equation}\label{App-A5}
  \langle\overline{D(t)}\rangle=\frac{1}{T}\int_0^T \langle D(t)\rangle dt.
\end{equation}
Substituting Eqs. \eqref{App-A4} and \eqref{App-A5} into Eq. \eqref{EB-Dt}, we obtain the result Eq. \eqref{EB-case2} in the main text.

\section{EB parameters for $D(t)=B^2(t)$}\label{App2}
For the diffusing diffusivity $D(t)$ being the square of Brownian motion, it holds that $\langle D(t)\rangle=\langle B^2(t)\rangle=t$. Combining it with Eq. \eqref{App-A5} yields that
\begin{equation}
  \langle\overline{D(t)}\rangle= \frac{T}{2}.
\end{equation}
On the other hand, by assuming $t_1<t_2$ and using the independent and stationary increments of Brownian motion $B(t)$, we have
\begin{equation}
  \begin{split}
    \langle B^2(t_1)B^2(t_2)\rangle&=\langle B^2(t_1)[B(t_2)-B(t_1)+B(t_1)]^2\rangle  \\
    &=\langle B^4(t_1)+ B^2(t_1)[B(t_2)-B(t_1)]^2 \rangle \\
     &~~~~+\langle2B^3(t_1)[B(t_2)-B(t_1)]\rangle  \\
     &=2t_1^2+t_1t_2.
  \end{split}
\end{equation}
Substituting this result into the expression of $\langle\overline{D(t)}^2\rangle$, we obtain
\begin{equation}
  \begin{split}
    \langle\overline{D(t)}^2\rangle&=\frac{2}{T^2}\int_0^Tdt_2\int_0^{t_2}dt_1\langle B^2(t_1)B^2(t_2)\rangle  \\
    &=\frac{7}{12}T^2.
  \end{split}
\end{equation}
Therefore, the EB parameter is
\begin{equation}
  \lim_{T\rightarrow\infty}\textrm{EB}=\frac{4}{3}.
\end{equation}


\bibliography{Reference}

\end{document}